\documentclass[showpacs,twocolumn,prb,floatfix,amssymb,aps]{revtex4}
\usepackage{graphicx}
\usepackage{natbib}

\begin{document}

\hsize\textwidth\columnwidth\hsize\csname@twocolumnfalse\endcsname

\title{Fractional Quantum Hall States at 1/3 and 5/2 Filling: the Density-Matrix Renormalization Group Calculations}
\author{Jize Zhao and D. N. Sheng}
\affiliation{Department of Physics and Astronomy, California State University, Northridge, California 91330, USA}
\author{F. D. M. Haldane}
\affiliation{Department of Physics, Princeton University, Princeton, NJ 08544}
\date{\today}
\begin{abstract}
In this paper, the density-matrix renormalization group 
method is employed to investigate
the fractional quantum Hall effect at filling fractions 
$\nu=1/3$ and $5/2$. 
We first present benchmark results at both filling fractions for
large system sizes to show  the  accuracy 
as well as the capability of the  numerical algorithm.
Furthermore, we show that by keeping a large number of basis states,  
one can also obtain accurate entanglement spectrum at $\nu=5/2$ for
large system with electron number up to $N_e=34$, much larger than
systems  previously studied. 
Based on a finite-size scaling analysis, we demonstrate that
the entanglement gap defined by Li and Haldane [\onlinecite{LI1}] 
is finite in the thermodynamic limit, 
which characterizes the  topological order of  the FQHE state.
\end{abstract}

\maketitle
\section{INTRODUCTION}
In condensed matter physics, one of the major challenges is to understand 
the strong correlation effect in interacting  electron
systems. The fractional quantum Hall effect (FQHE) 
systems\cite{TSUI1} are primary examples, where  new quantum phases emerge
with fractional quasiparticle excitations\cite{LAUGHLIN1}
resulting from such effect. 
Theoretical understanding\cite{LAUGHLIN1,HALDANE1,HALPERIN1,YOSHIOKA1,JAIN1}
of the odd denominator
FQHE states in the lowest Landau level $(n=0)$ has been fully 
developed soon after the experimental discovery\cite{TSUI1}.
The even denominator\cite{WILLETT1} quantum Hall states in second Landau level $(n=1)$ 
appear to demonstrate complex nature beyond the understanding of a unified theory,
which are under  intensive studies\cite{HALDANE2,MOORE1,WEN1,FRADKIN1,READ1,XIA1,SARMA1,NAYAK1,LEVIN0,FISHER1,PETERSON1,WANG1}.
 
Theoretical approaches\cite{JAIN1,MOORE1,WEN1,FRADKIN1,NAYAK1} 
have predicted a variety of possible 
candidate states for even-denominator quantum Hall systems, 
some of which are exotic in nature with quasiparticles obeying
non-Abelian statistics.  
However, very often such theories 
cannot provide  solutions to  microscopic theoretical models describing
realistic electron systems.
In this aspect, computational studies  have  played an important
role to determine the quantum state for  such systems.
Pioneer works have been done
using  exact diagonalization (ED) method 
to  establish the
nearly perfect overlap  of 
the Laughlin  wave function   with
the exact ground state wavefunction of 
small electron systems  and
provide microscopic understanding of the nature of the FQHE
\cite{YOSHIOKA1,HALDANE1,HALDANE2}.   
In recent years, ED has been widely used to study FQHE for pure 
systems\cite{FANO1,AMBRUMENIL1,MORF0,MORF1,WANG1} 
as well as disordered systems\cite{WAN1} to probe the nature of various
quantum phases and transitions. It has been recently established that
ED can also be used to identify the
topological nature of a quantum phase based on entanglement entropy
and the entanglement spectrum
of the quantum systems\cite{LI1,HAQUE1,LAEUCHLI1,ZOZULYA1,BERNEVIG1}.
However, since the Hilbert space increases exponentially with the system size, 
ED is limited to systems with a small electron number 
(typically restricted to electron number $N_e=14\sim 20$ depending on
the Landau level filling number).
This limitation becomes a severe problem when the finite-size effect is strong,
which is usually the case for quantum states in the higher
Landau levels or with even denominator filling fractions.

The density-matrix renormalization group (DMRG) 
developed by White is a powerful method 
for studying  interacting systems with  
accuracy controlled by the number of states kept in  DMRG
blocks\cite{WHITE1, SCHOLLWOCK1}. 
It has been widely applied to quasi-one-dimensional systems, providing 
essentially exact results for spin or electron systems. 
However, its application to 
two-dimensional systems remains to be quite challenging. 
It is generally believed that the number of states
desired to be kept in each block should grow exponentially  
with the increase of the system width
to catch up the entanglement entropy between two coupled 
blocks\cite{VERSTRAETE1}. 
Nevertheless, FQHE systems can be modeled 
as  one dimensional system with long-range Coulomb interaction,
which may become accessible using current computational power.
Shibata and Yoshioka made the first attempt to develop
DMRG algorithm to study quantum Hall systems in torus 
geometry\cite{SHIBATA1} about ten years ago.
Various fillings have been studied\cite{SHIBATA2} by keeping  hundreds of 
states obtaining useful information of larger systems at integer
fillings or for compressible states.
Recently, a great progress has been made by Feiguin $\it {et. al}$\cite{FEIGUIN1, FEIGUIN2} 
developing new DMRG algorithm keeping up to 5,000 states 
in their work. The hard problems of the incompressible
FQHE at $\nu=1/3$ and $5/2$ have been extensively
studied, providing convincing  high accuracy
ground state and excited state results for larger system sizes up to
20 and 26 electrons at $\nu=1/3$ and $5/2$, respectively.
Moreover, DMRG has also been applied to study bosonic quantum Hall 
effect\cite{KOVRIZHIN1}.

In this paper,  we study FQHE systems based on our newly developed
DMRG code, which substantially improves the accuracy of the
results through keeping more states and managing DMRG process with
higher efficiency.
The benchmark results and error analysis are presented for $\nu=1/3$ and
$5/2$. For the same system sizes studied before,
we  reduce the error substantially especially for $5/2$
systems, while we access systems
with more electrons with high accuracy.
We also obtain accurate entanglement spectra for $\nu=5/2$ systems,
which identify the topological order at larger system sizes.
The entanglement gap for $\nu=5/2$ FQHE
first revealed  by Li and Haldane\cite{LI1} based on  ED calculation, 
remains finite in the thermodynamic limit established through 
finite-size scaling analysis.

\section{HAMILTONIAN AND METHOD}
In this paper, we study the FQHE in the spherical geometry\cite{HALDANE1}, 
where electrons are confined on the surface of a sphere with radius $R$.
The total magnetic flux through the spherical surface $4\pi R^2 B$ are
quantized to be an integer $2S$ multiple of the flux quanta.
Assuming that electrons are polarized by the magnetic field
and neglecting the Landau level mixing,
the Hamiltonian  in the spherical geometry can be written as 
\begin{eqnarray}
\mathcal{H}=\frac{1}{2}\sum_{m_1,m_2,m_3,m_4}\langle m_1m_2|V|m_3m_4\rangle a^{\dagger}_{m_1}
            a^{\dagger}_{m_2}a_{m_3}a_{m_4} 
\label{HAM}
\end{eqnarray}
where the $m_i$ is the z-component of the angular momentum
and $m_i=-L,-L+1,\cdots,L$, with $L=S+n$ being the total angular momentum
and $n$ is the Landau level index.
$V$ is the Coulomb interaction between electrons in units of
 $\frac{e^2}{l_0}$, with  
$l_0=\sqrt{\hbar c/eB}$ being the magnetic length. In the next sections, we also use the renormalized 
magnetic length $l_0^{(\infty)}$ to rescale the energy unit if it is mentioned explicitly\cite{AMBRUMENIL1,MORF1}.  
$a_m (a^{\dagger}_m)$ is the
annihilation (creation) operator at the  orbital $m$. 
The problem can be naturally studied
using the DMRG method in momentum space\cite{XIANG1}
as nonzero matrix elements in the Eq. (\ref{HAM})
only exist between orbitals   
satisfying the angular momentum conservation relation $m_1+m_2=m_3+m_4$.

In our DMRG process, we arrange the $2L+1$ orbitals into a one-dimensional chain, 
corresponding to states with z-component of the angular momentum $m$ 
taking different values $L,L-1,\cdots,-L+1,-L$ from left to right. 
In the initial process, we start from 
a small system with two blocks (each block has only one orbital)
and two single orbitals in the middle
forming a configuration  $B_L\bullet\bullet B_R$. 
We write the Hamiltonian (\ref{HAM}) terms using  the product of 
block operators and
single site operators\cite{WHITE1,XIANG1,SCHOLLWOCK1}  
and diagonalize the Hamiltonian to obtain the ground state. The
reduced density matrix is formed and diagonalized by following the standard 
DMRG\cite{WHITE1} procedure.  
A new block $B_{L}$ ($B_R$) is formed by including the single site into the 
current block.
This procedure is repeated until the 1D chain grows into the
desired system size with 2L+1 orbitals. From this point,
the finite-size sweeping algorithm is employed until 
we obtain the converged results.
\begin{figure}
\includegraphics[width=7.0cm,clip]{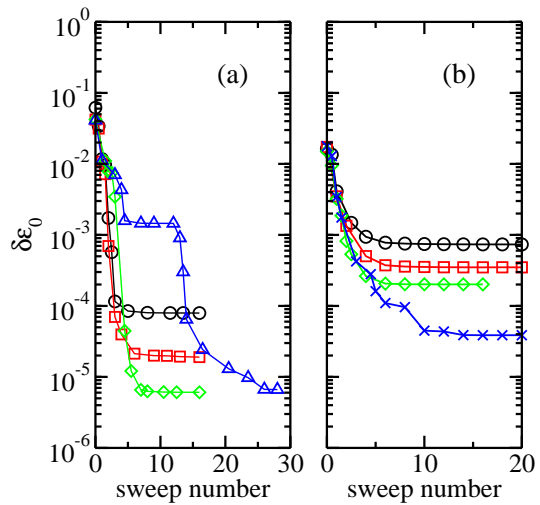}
\caption{(color online)  Error in the ground state energy per electron $\delta \epsilon_0$
is shown at (a) $\nu=1/3$ and $N_e=20$, with 1,600 ($\bigcirc$), 3,200 ($\square$), and 5,000 
($\Diamond$) states kept, respectively. The triangle data(obtained by keeping 5,000 states) illustrate a situation where 
DRMG runs into a local minimum during initial 12 sweeps, see text for details.  
The reference ground state energy is -0.4210509, obtained by DMRG with at most 24,000 states kept.
(b) $\nu=5/2$ and $N_e=24$ with 1,600($\bigcirc$), 3,200($\square$), and 5,000($\Diamond$) states kept  
, respectively. The data showed by $\times$ are obtained 
by keeping 8,000, 10,000 and 14,000 states for each 4 sweeps starting
from the first sweep, and 16,000 states for the rest of sweeps. 
The reference ground state energy is -0.3876150, obtained by 
DMRG with at most 30,000 states kept. 
}
\label{BENCHMARK}
\end{figure}

Using our momentum space DMRG, 
the Hamiltonian (\ref{HAM}) is diagonalized in the sector with fixed 
electron number $N_e$ and the total angular momentum z-component $L_z$.
In the initial process of the calculation, only a small
fraction of the momentum sectors in each
block contributes significantly to the ground state wavefunction. 
States in other sectors, which have
zero eigenvalues in reduced density matrix,
are being discarded. However, these discarded
states may become also important to the ground state wavefunction
in the later stage when system grows to the full length\cite{XIANG1}.
To overcome this
problem, we need to keep additional sectors from very
beginning. For this purpose, we set a minimum
number of sectors $N_{sec}^{min}$, which in practice is about $ 3 \sim 5$ times
larger than the number of sectors after convergence.
In each of selected sectors, we keep at least two states with largest 
eigenvalues in the sector.
The remaining states are selected following the standard DMRG procedure
to minimize the truncation error.
In the sweep process, $N_{sec}^{min}$ is gradually decreased as a function of sweep number.

In Fig. \ref{BENCHMARK}(a), we show the  error of the obtained
ground state energy per electron at $\nu=1/3$ for different number of
states kept in each  block
with respect to the fully converged reference energy obtained
by keeping much more states.
A reasonable accuracy around $10^{-4}$ is reached with the 
electron number $N_e=20$ (bigger than the largest ED size by six electrons)
by keeping only $1,600$ states, while
we achieve the accuracy of $6\times 10^{-6}$ by keeping $5,000$ states. 
However, the convergence  becomes more difficult for  $\nu=5/2$ system, 
as we show in Fig. \ref{BENCHMARK}(b). 
For $N_e=24$ (bigger than the largest ED size by four electrons), 
with $m=5,000$ states kept, the accuracy we can achieve
is about $2\times 10^{-4}$, comparable with the results
obtained by Feiguin et al\cite{FEIGUIN1}. 
Further increasing the number of states kept to $16,000$, 
we are able to reduce the error by a factor of five.

Controlling $N_{sec}^{min}$ also allows us to 
overcome the  local minimum trapping, which
could trap the DMRG obtained state  in an excited state.
As we show in Fig. \ref{BENCHMARK}(a) by triangles,  
the energy is pinned to a local minimum from the fifth sweep to twelfth 
sweeps, and the error is much larger than 
the data showed by $\Diamond$ starting from a different initial state. 
To overcome the problem, we simply 
increase $N_{sec}^{min}$ at the twelfth sweep to allow a larger number of sectors 
to get into the Hilbert space. 
These additional sectors bring in significant quantum
fluctuations and eventually get
the state out of the local minimum.

\section{RESULTS}
\subsection{Ground state energies at $\nu=1/3$ and $5/2$}
To demonstrate the accuracy of our DMRG calculations, 
we have obtained the ground state energy for system  up to 24 electrons
at $\nu=1/3$ by keeping up to
$20,000$ states,  which leads 
to a truncation error smaller than $10^{-11}$ in the final sweep.
The maximum dimension of the Hilbert
space diagonalized is of the order of $2\times 10^7$.   
\begin{figure}[ht]
\includegraphics[width=7.0cm,angle=0,clip]{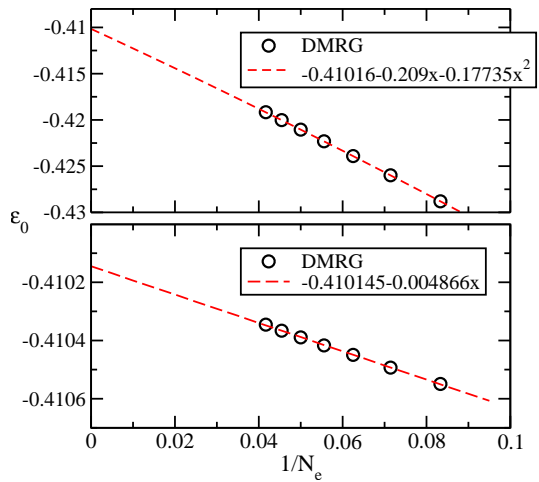}
\caption{(color online) (a) Finite-size scaling of ground state energy per electron 
at $\nu=1/3$ with a quadratic extrapolation.
         (b) Finite-size scaling of ground state energy per electron at $\nu=1/3$ 
with a linear extrapolation. Energy unit is 
renormalized by $l^{(\infty)}_0$. 
The error bars in both figures are much smaller than the size of the symbols.}
\label{FIG2}
\end{figure}
In Fig. \ref{FIG2}, we show the ground state energy as a function of $1/N_e$ with $2L=3(N_e-S_0)$, 
where a shift of $S_0=1$ has been used. In the upper panel numerical 
data are extrapolated to thermodynamic limit by a quadratic function of $1/N_e$, which gives the ground state 
energy per electron $\epsilon_0=-0.41016(2)$. In the lower panel we rescale 
the energy by the renormalized magnetic length\cite{AMBRUMENIL1,MORF1}  
and extrapolate the numerical data linearly, leading to $\epsilon_0=-0.410145(15)$  demonstrating   consistency between two extrapolating methods.
While our results are  essentially in agreement with 
previous results recently obtained by Feiguin et al\cite{FEIGUIN1} 
using DMRG method, 
our accuracy is improved by
keeping much more states, which opens opportunity 
for studying larger systems. 
Now we turn to the study of the FQHE at $\nu=5/2$. 
We calculate the ground state energy 
up to  $N_e=34$ electrons, with at 
most 24,000 states kept. The maximum dimension of the space we diagonalize is 
around  $3.5\times 10^{7}$. The truncation
error is of the order of $10^{-7}$ indicating larger error in
DMRG comparing to $\nu=1/3$ case due to larger entanglment entropy in the ground state 
as well as smaller gap separating the excited
states from the ground state. The data at the largest system size were obtained
within three weeks on one 12 cores  Xeon server.  
\begin{figure}[ht]
\includegraphics[width=7.0cm,angle=0,clip]{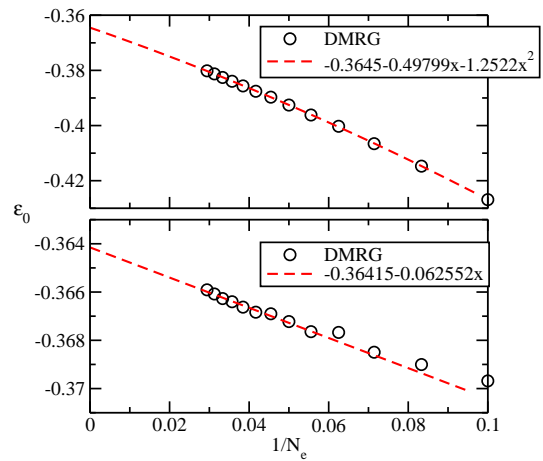}
\caption{(color online) (a) Finite-size scaling of the ground state energy at $\nu=5/2$ with a quadratic fitting.
         (b) Finite-size scaling of the ground state energy at $\nu=5/2$ with a linear fitting. Energy unit is 
renormalized by $l_0^{(\infty)}$ and only $N_e\ge 18$ is used in view of the strong finite-size oscillation.} 
\label{FIG3}
\end{figure}
In upper panel of Fig. \ref{FIG3}, we show the ground state energy $\epsilon_0$ as a function of $1/N_e$. 
With a quadratic fitting, the extrapolated value in the thermodynamic limit 
is $-0.3645(6)$. In the lower panel,
we rescale the ground state energy by the renormalized magnetic length $l^{(\infty)}_0$ and extrapolated ground state energy 
is $-0.36415(45)$ by a linear function. 
We further calculate the excitation gaps at $\nu=5/2$ up to $N_e=26$, 
including neutral exciton gap($\Delta^{exc}$) and the charged excitation gap($\Delta$), following the definitions in
Ref. \onlinecite{MORF1}. Some of the data are present in Table \ref{TABLE1} as a function of $N_e$, 
where the excited states ``aliased'' to other  quantum Hall states are excluded\cite{MORF0,MORF1}.
The estimated gaps in the thermodynamic
limit are $\Delta^{exc}=0.032\pm 0.004$ and $\Delta=0.029\pm 0.003$, 
in agreement with early result,
although we observe that the strong finite-size oscillation 
persists even for our larger system size data. 
\begin{table}
\caption{Excitation gaps at $\nu=5/2$ as a function of $N_e$. Some data for exciton gaps are not shown here.}
\begin{ruledtabular}
\begin{tabular}{c|cccccc}
$N_e$            & 10      & 14       & 18      & 22  & 26 \\
\hline
$\Delta^{exc}$   & 0.03905 & 0.03981  &  0.03751  & 0.03823 & 0.03982 \\
$\Delta$         & 0.04518 & 0.03934  &  0.03619 & 0.03756 & 0.03482
\end{tabular}
\end{ruledtabular}
\label{TABLE1}
\end{table}

\subsection{Entanglement spectrum and entanglement gaps}
\begin{figure}
\includegraphics[width=7cm,clip]{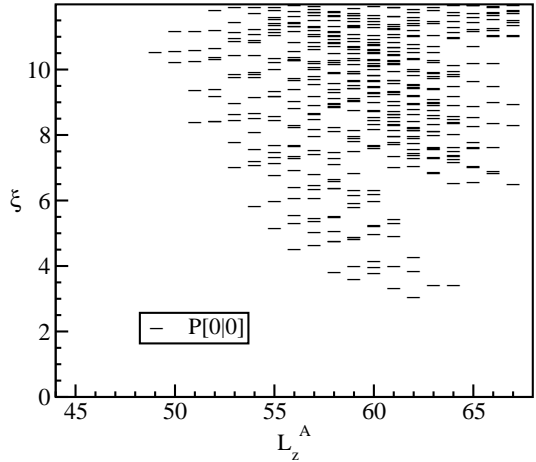}
\caption{The low-lying entanglement spectrum of $N_e=16$ for the partition $P[0|0]$ obtained by
DMRG are shown.}
\label{ES-16}
\end{figure}
By dividing $2L+1$ orbitals into two parts, $A$ and $B$, the ground state can be written, 
according to Schmidt decomposition, as
\begin{eqnarray}
|\psi\rangle = \sum_i e^{-\xi_i/2} |\psi^i_A\rangle \otimes |\psi^i_B\rangle, \label{ENG}
\end{eqnarray}
where the singular eigenvalues $\exp(-\xi_i/2)$ obtained from diagonalizing 
the reduced density matrix
define the entanglement spectrum ${\xi_i}$. The  
$|\psi^i_A\rangle$ and $|\psi^i_B\rangle$ are the orthogonal
basis states of part $A$ and $B$, respectively.
In the pioneer work of Li and Haldane\cite{LI1}, 
the entanglement spectrum for 
system at $\nu=5/2$ has been analyzed based on exact diagonalization calculation
up to $N_e=16$ electrons. The obtained
entanglement spectrum show the same structure as the conformal
field theory (CFT) for the Moore-Read state, below an entanglement 
gap while the non-CFT type of spectrum exist above the gap. 
Therefore, the entanglement spectrum reveals more information than 
the entanglement entropy\cite{HAQUE1,LAEUCHLI1}. 
Although the entanglement spectrum and
entropy are  naturally obtained in DMRG, they are much harder
to  converge\cite{SHENG2}
than the ground state energy.  
In the following, we demonstrate the success of our DMRG 
calculations in this aspect by keeping up to 24,000 states.

\begin{figure}[ht]
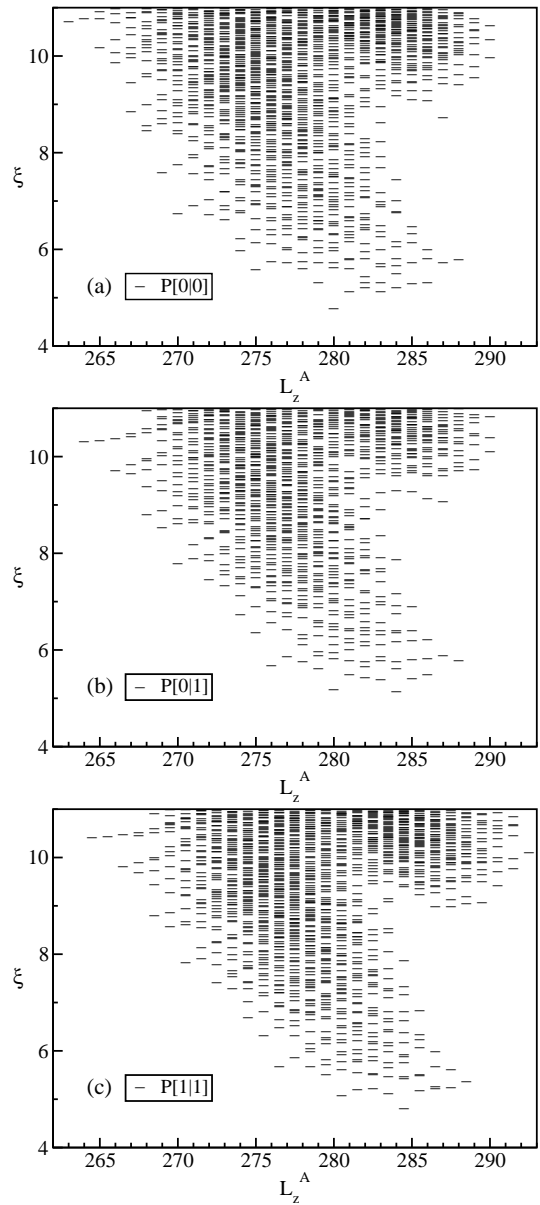

\vspace{0.5cm}
\includegraphics[width=7.0cm,clip]{fig5.eps}
\includegraphics[width=7.0cm,clip]{fig6.eps}
\includegraphics[width=7.0cm,clip]{fig7.eps}
\caption{The low-lying entanglement spectra of $N_e=34$ are shown for three
partitions $P[0|0]$, $P[0|1]$ and $P[1|1]$.}
\label{ES}
\end{figure}
We note that our two parts of system 
 $A$ and $B$ used for Eq. (2) include one block and one site defined
for the DMRG process, respectively, 
thus, the quantum numbers from both parts satisfy the equations
 $N_e^A+N_e^B=N_e$ and $L_z^A+L_z^B=0$ (the ground state for 5/2
FQHE  has total $L_z=0$). 
The entanglement spectrum can be labeled by $N_e^A$ and $L_z^A$ in part $A$. 
We recall that\cite{BERNEVIG1,LI1} for the ground state at $\nu=5/2$, the
highest-density ``MR root configuration'' has a pattern of 
``11001100$\dots$110011'' corresponding to the ``generalized Pauli principle'' that 
no group of 4 consecutive orbits contains more that 2 particles.
Consequently, there are three distinct ways of partitioning the orbits
as between two zeros, between zero and one, and between two ones.
These three different partitions of the root configuration, 
 identified as $P[0|0]$, $P[0|1]$, and 
$P[1|1]$ can be  easily obtained in the  sweeping process\cite{WHITE1}
during the DMRG calculations.

To check the accuracy we compare the entanglement spectrum obtained by 
DMRG with that obtained by ED for $N_e=16$ for the partition $P[0|0]$, as 
shown in Fig. \ref{ES-16}. DMRG results precisely  reproduce
ED results\cite{LI1}.     
Next, we turn to consider larger size. In Fig. \ref{ES}, we show the 
entanglement spectra of three partitions for $N_e=34$. 
The low-lying spectra have the identical counting structure viewing
as a function of the
$\Delta L=L_{z, max}^A-L_z^A$ (here the $L_{z,max}^A=288$ for
the partition $P[0|0]$) comparing to the results presented
 by Li and Haldane [\onlinecite{LI1}], establishing the
identical topological order as in the Moore-Read state
at this larger system size.
Furthermore, the same spectrum structure 
has also been obtained for 
a wide range of even electron numbers  $8\leq N_e \leq 34$.

\begin{figure}[ht]
\includegraphics[width=6.5cm,clip]{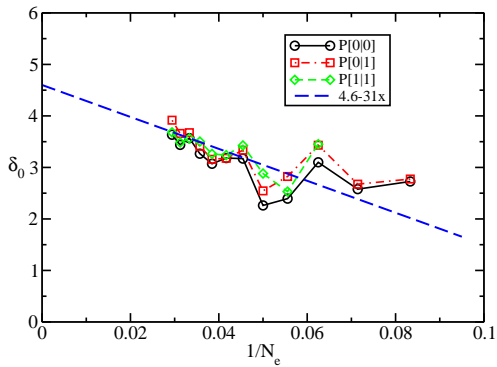}
\caption{(color online). Entanglement gap at $\Delta L=0$ vs $1/N_e$
for three different partitions. Dashed blue line is a linear 
fit for $N_e\ge 16$.}
\label{EGAP}
\end{figure}
Finite-size calculation shows that the CFT part of entanglement spectrum is 
protected by an entanglement gap from the non-CFT part and,
the entanglement gap plays the key role of measuring the robustness of topologically ordered state\cite{LI1}.  
In literature, the entanglement gap is obtained by ED and limited to a small size\cite{LI1, THOMALE1}
and whether it persists in the thermodynamic limit remains an open question. 
Thus, identifying whether the entanglement gap persist in the thermodynamic limit 
is particularly important for the establishment of the topologically 
ordered state in such systems. This is different from the  finite-size  wavefunction 
overlaps between the wavefunction of a realistic interacting 
system and model wave-functions (Laughlin or Moore-Read wavefunctions) 
as such overlaps have  to go to zero
with the increase of the system size related to the presence of the
generic entanglement spectrum in the  system. 
Here we have obtained the entanglement spectrum gap for a relatively
large range of the system sizes, which 
allows us to extrapolate the gap to the thermodynamic limit. 
In Fig. \ref{EGAP}, we show the entanglement gap $\delta_0$ at $\Delta L=0$ as a function of $1/N_e$
for three partitions. Entanglement gap oscillate strongly for small sizes, and the oscillation magnitude 
becomes smaller as $N_e$ increases. Moreover, the lines for the three partitions almost fall into one curve. 
We fit the entanglement gap as a linear function of $1/N_e$ and neglect higher order corrections
due to the oscillation of the data and limitation of the size.
It yields $\delta_0=4.6\pm 0.6$ in the thermodynamic limit
and thus our work confirms the conjecture of a finite entanglement gap made by Li and Haldane \cite{LI1}.

\section{CONCLUSIONS}  
We have presented  systematic numerical results
obtained by a newly developed   DMRG program for  FQHE systems.
We have substantially  improved the DMRG algorithm and 
obtained accurate results for ground state energy and excitation
gap for larger systems than previous works 
by ED and DMRG at $\nu=1/3$ and  $5/2$.
In particular,  we demonstrate the robustness of the low-lying
CFT entanglement spectrum for large system size with $N_e=34$ and the
finite entanglement gap in the thermodynamic limit at $\nu=5/2$
based on finite-size scaling.

J. Zhao is grateful to H. T. Lu and C. T. Shi for helpful discussion.  
This work is  supported by
 US DOE Office of Basic Energy Sciences under grant
DE-FG02-06ER46305 (DNS), the NSF grants DMR-0611562, 
 DMR-0906816 (JZ), and  MRSEC Grant  DMR-0819860 (FDMH).

\vfill
\end{document}